\documentclass{article}
\usepackage{spconf,amsmath,graphicx,bm}
\usepackage{subcaption}
\usepackage{tikz}
\usepackage{multirow}
\usetikzlibrary{positioning}
\usepackage{todonotes}
\usepackage{url}
\usepackage{booktabs}
\usepackage{amsfonts}

\title{ENERGY-BASED MODELS FOR SPEECH SYNTHESIS}
\name{Wanli~Sun, Zehai~Tu, Anton~Ragni}
\address{Department of Computer Science, University of Sheffield, Sheffield, UK\\
\small{\tt \{wsun20, ztu3, a.ragni\}@sheffield.ac.uk}}

\ninept
\begin{document}
\maketitle
\begin{abstract}
Recently there has been a lot of interest in non-autoregressive (non-AR) models for speech synthesis, such as FastSpeech 2 and diffusion models. Unlike AR models, these models do not have autoregressive dependencies among outputs which makes inference efficient. This paper expands the range of available non-AR models with another member called energy-based models (EBMs). The paper describes how noise contrastive estimation, which relies on the comparison between positive and negative samples, can be used to train EBMs. It proposes a number of strategies for generating effective negative samples, including using high-performing AR models. It also describes how sampling from EBMs can be performed using Langevin Markov Chain Monte-Carlo (MCMC). The use of Langevin MCMC enables to draw connections between EBMs and currently popular diffusion models. Experiments on LJSpeech dataset show that the proposed approach offers improvements over Tacotron 2. 

\end{abstract}
\section{Introduction}
\label{sec:intro}

Neural network based synthesis have made impressive improvements over statistical speech synthesis. However, these deep learning based text-to-speech (TTS) approaches often feature inconsistencies as did statistical approaches. 
For example, auto-regressive (AR) models, such as Tacotron 2 \cite{shen2018natural}, Transformer-TTS \cite{li2019neural},  
are almost exclusively trained using teacher forcing~\cite{williams1989learning}, where reference rather than predicted values are fed back into the generative process. Such a mismatch between training and inference causes inconsistency called \textit{exposure bias}~\cite{Ranzato2015SequenceLT}, which may lead to poor generated speech quality (e.g. repetition, skipping, and long pauses \cite{shen2020non}). 
So far there have been a few attempts to alleviate exposure bias, such as, scheduled sampling \cite{bengio2015scheduled}
and attention mechanisms \cite{He2019RobustSA}. However, their effective application is complicated due to a number of ``training hacks" employed to ensure stable learning ~\cite{battenberg2020location}.

Recently, there has been interest in non-AR models, such as FastSpeech 2 \cite{ren2020fastspeech} and diffusion models \cite{popov2021grad}. 
These models generally do not use teacher forcing as a part of their training and hence should be free of the aforementioned inconsistencies. This paper describes another class of non-AR models called energy-based models\,(EBMs) \cite{lecun2006tutorial}, which, as will be shown later, have connections to currently popular diffusion models. Given a text, an EBM defines an energy-function over all possible spoken realisations. 
Although it is possible to formulate the conditional probability distribution of speech given text for EBMs, the intractable normalisation term would make training and inference approaches relying on the probability distribution infeasible. 

Instead, training of EBMs can be performed using noise contrastive estimation (NCE), which compares speech data (positive examples), which is assumed to represent high quality speech, and imperfect speech data (negative examples). The nature of imperfection, or negative examples, is crucial when training EBMs. This paper describes a number of effective strategies to generate negative examples, including by means of existing TTS models. 
Inference with EBMs can be performed using Langevin Markov Chain Monte-Carlo (MCMC) \cite{parisi1981correlation, grenander1994representations}. Given that a similar iterative algorithm is often used with diffusion models (e.g., Grad-TTS \cite{popov2021grad}), this paper discusses connections between EBMs and diffusion models.

This paper makes the following specific contributions:
\begin{enumerate}
\item first energy-based text-to-speech model;
\item a range of methods for generating effective negative samples to use in NCE and elsewhere;
\item link between diffusion models and energy-based models.
\end{enumerate}

The rest of this paper is organized as follows. Section \ref{sec:R-EBMs} describes energy-based models (EBM), which includes inference, training and negative sampling methods. Section \ref{sec:prior} relates EBMs to filtering approaches and diffusion models. Experimental results and discussion are presented in Section \ref{section-experiments}. Conclusions drawn from this work and future research directions are presented in Section \ref{sec:conclusion}.

\section{Energy-based models}
\label{sec:R-EBMs}
Given a text sequence ${\bm x}$, an energy-based model (EBM) of speech feature sequences ${\bm Y}$ (e.g. log-Mel spectrograms) can be defined by\footnote{An alternative formulation would involve parameterising the gradient of energy instead. Such an approach is possible due to existence of inference and training approaches that rely only on the gradient of energy. }
\begin{equation}
p_{\bm \theta}(\bm{Y} | \bm{x})=\frac{1}{Z_{\bm \theta}(\bm {x})} \exp \left(-E_{\bm{\theta}}(\bm{x}, \bm{Y})\right),
\label{EBM}
\end{equation}
where ${\bm\theta}$ are model parameters, $E_{\bm{\theta}}(\bm{x}, \bm{Y})$ is an energy between text and speech, $Z_{\bm \theta}(\bm {x})$ is a normalisation term. Unlike speech signal energies commonly used in models like FastSpeech 2, EBM energies $E_{\bm{\theta}}(\bm{x}, \bm{Y})$ reflect the correspondence between text ${\bm x}$ and speech ${\bm Y}$ pairs. Better matching pairs are expected to yield lower energies and {\em vice versa}. The normalising term $Z_{\bm \theta}(\bm {x})$ is intractable to compute exactly. Thus, only certain inference and parameter estimation approaches can be used for EBMs. 

\subsection{Inference}
For tasks where outputs are represented by discrete tokens ({\em e.g.} characters or words), such as text generation \cite{bakhtin2021residual} and speech recognition \cite{Li2021ResidualEM}, EBMs are commonly used to rerank hypotheses generated during beam search. In contrast, for tasks where outputs are represented by continuous variables, such as speech synthesis, EBMs can be used for updating hypotheses themselves. This can be done using Langevin Markov Chain Monte-Carlo (MCMC). The Langevin MCMC is an iterative process where, given an initial hypothesis ${\bm Y}^{(0)}$, the next hypothesis is obtained by
\begin{equation}
{\bm Y}^{(N+1)} \!\leftarrow\! {\bm Y}^{(N)} \!-\! \lambda \left.\nabla_{\bm Y}E_{\bm{\theta}}(\bm{x},  \bm{Y})\right|_{{\bm Y}={\bm Y}^{(N)}} + \sqrt{2\lambda} {\bm Z}^{(N)},
\label{langevin_mcmc}
\end{equation}
where ${\bm Z}^{(N)} \sim {\mathcal N}({\bm 0},\mu{\bm I})$, $\lambda$ is an updating rate and $\mu$ is commonly set to 1.
The need to specify initial hypotheses ${\bm Y}^{(0)}$ offers a number of interesting options. In the standard Langevin MCMC initial hypotheses are drawn from a simple prior distribution, such as Gaussian \cite{huszar2015not}. However, more informative priors, such as high-performing TTS models (e.g. Tacotron 2 and FastSpeech 2), can also be explored.

\subsection{Training}
\label{sec:training} 
Since the normalising factor ${Z_{\bm\theta}(\bm{x})}$ is intractable, approaches relying on $p_{\bm \theta}(\bm Y | \bm{x})$ can not be used with EBMs. Furthermore, popular gradient-based MCMC approaches \cite{ du2019implicit} can not be applied with discrete input, as in this work, and Gibbs sampling ~\cite{welling2004exponential} would be too computationally expensive. Fortunately, noise contrastive estimation (NCE) \cite{Ma2018NoiseCE} provides a feasible solution to optimize energy functions. The NCE loss function for EBMs in eq.~\eqref{EBM} is given by 
\begin{equation}
    \begin{split}
         \mathcal{L}_{\bm\theta}({\bm x},{\bm Y}^{+},{\bm Y}^{-}) = &-\log\left( \frac{1}{1+\exp \left(E_{\boldsymbol{\theta}}(\bm{x}, \bm{Y}^+)\right)}\right) \\
        &- \log \left(\frac{1}{1+\exp \left(-E_{\bm{\theta}}(\bm{x},  \bm{Y}^-)\right)}\right)
    \end{split}
    \label{loss function}
\end{equation}
where ${{\bm Y}^{+}}$ are called positive samples and ${{\bm Y}^{-}}$ are called negative samples. According to eq.~\eqref{loss function}, energy functions are optimal when high energy is assigned to negatives and low energy is assigned to positives. Once trained, energy functions can be used for ranking hypotheses generated by other models or inferring hypotheses using the Langevin MCMC in eq.~\eqref{langevin_mcmc}.

Positive samples in NCE are usually represented by reference sequences. On the other hand, negative samples need to be designed. In text generation~\cite{bakhtin2021residual} it is argued that negative examples with poor quality make the task of learning the energy function easier which leads to poor quality energy functions. This work proposes using pre-trained TTS models to generate high-quality negative examples. Note that when a pre-trained model is used as a part of training process then it is also possible to adopt it for initialising the Langevin MCMC in eq.~\ref{langevin_mcmc}, which is expected to speed up inference and lead to higher quality hypotheses. The simplest method to generate negative samples from pre-trained TTS models would use hypotheses generated by those models directly. Such approach may fail to work due to high level of similarity between high-quality hypotheses and reference sequences. Other possible options include applying random masking (RM) and SpecAugment~\cite{park2019specaugment} ({\em i.e.} time masking (TM), frequency masking (FM) and time warping (TW)) to those hypotheses. The TM and FM methods can be seen as specific cases of the RM method and may prove less effective. For example, the FM may drastically affect pitch information by masking a whole frequency range. The TW method is also expected to face challenges as utterance-wide shortening/elongation of all sounds may be hard to separate from reference sequences. 

\begin{figure}[t]
  \centering
  \includegraphics[width=\linewidth]{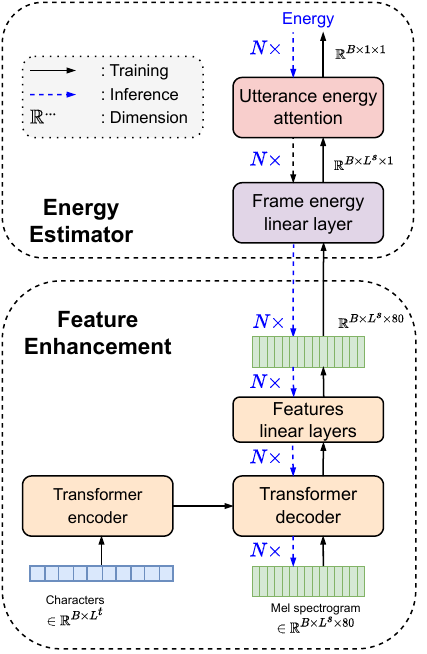}
  \caption{Architecture of EBMs examined in this work (inspired by Transformer TTS)}
  \label{fig:structure}
\end{figure}
\subsection{Architecture}
The architecture of EBM explored in this work is inspired by Transformer TTS \cite{li2019neural} and is shown in Figure~\ref{fig:structure}. The EBM in Fig.~\ref{fig:structure} consists of two blocks: energy estimator (top) and feature enhancement (bottom). The goal of the energy estimator is to derive utterance-level EBM energies $E_{\bm\theta}({\bm x},{\bm Y})$ from the output of the feature enhancement block. This work assumes that these energies can be derived from frame-level EBM energies. As shown in Fig.~\ref{fig:structure}, the energy estimator consists of two key elements: frame-level EBM energy estimation and frame-level EBM energy weighting. The latter element is motivated by an intuition that frame-level EBM energies are unlikely to make equally important contributions. For example, speech and non-speech frames will likely make different contributions to the utterance-level EBM energy 

\begin{equation}
E_{\bm{\theta}}(\bm{x}, \bm{Y}) = \sum_{t=1}^{T} \alpha_{t} e_{t}
\label{sequence-level-energy}
\end{equation}
where ${\alpha}_{1:T}$ is the sequence of attention weights generated by the EBM energy weighting module, $e_{1:T}$ is the sequence of frame-level EBM energies and $T$ is the number of frames. The attention weights are derived from the frame-level EBM energies. The frame-level EBM energies ${e}_{1:T}$ are computed by
\begin{equation}
e_{t} = {\bm a}^{\top} {\bm g}_{t} + b
\end{equation}
where ${{\bm g}_{t}}$ is the output of the feature enhancement block, and $\bm a$ and $b$ are parameters of the frame energy module. The goal of the feature enhancement block is to enhance typically short-term spectral information available in standard speech features, such as log-Mel spectrograms, with more advanced acoustic and linguistic information. The estimator is a transformer-based~\cite{li2019neural} model using text as the input to encoder and and spectral features and the output of encoder as the input to decoder. The output of decoder after linear transformation, ${{\bm g}_{t}}$, is passed to the energy estimator block. Note that the decoder does not use masking to constrain the underlying attention mechanism from attending over previous spectral features, which makes ${{\bm g}_{t}}$ a function of entire text and spectral feature sequences.  

\section{RELATED WORK}
\label{sec:prior}
EBMs have been applied in a wider range of domains, e.g. natural language Processing \cite{bakhtin2021residual, clark2020pre} and automatice speech recognition \cite{Li2021ResidualEM}. The EBM proposed in this work can be related to a number of previously proposed approaches in TTS. The use of hypotheses generated by pre-trained TTS models as a part of training and inference allows to connect this EBM to post-filtering methods. Statistical post-filtering approaches, such as \cite{Takamichi2014APT}, aim to address over-smoothing in hypotheses generated by statistical speech synthesis models. However, these approaches suffer from difficulties in accurately modelling probability density functions of the underlying speech parameterisations. Recently, there has also been interest in deep learning based post-filtering approaches. In \cite{kaneko2017generative}, frequency band specific generative adversarial networks (GAN) were trained to improve the quality of hypotheses generated by deep learning based speech synthesis models. However, this approach assumes independence among frequency bands which may lead to suboptimal results. 

More recently, there has been a lot of interest in diffusion-based TTS models \cite{popov2021grad}, which have been extended to audio synthesis \cite{pascual2023full} and singing voice synthesis \cite{xue2022learn2sing}. In these models training (forward) and inference (reverse) processes iteratively build a connection between data and noise. Although seemingly different, such diffusion models and the EBM proposed in this work have clear connections. Consider, for example, the iterative inference process used by one of those diffusion models \cite{song2020score} 
\begin{equation}
{\bm Y}^{(N+1)} \!\leftarrow\! \frac{1}{\sqrt{1\!-\!\lambda_N}}({\bm Y}^{(N)} \!+\! \lambda_{N} S_{\bm{\theta}}(\bm{x},\bm{Y}^{(N)},{N})) + \sqrt{\lambda_{N}} {\bm Z}^{(N)},
\end{equation}
Compared to the Langevin MCMC in eq.~\eqref{langevin_mcmc} the key difference stems from modelling iteration, $N$, specific $S_{\bm{\theta}}(\bm{x},\bm{Y}^{(N)},{N})$ score (gradient of log-likelihood) rather than iteration independent $\nabla_{{\bm Y}^{(N)}}E_{\bm{\theta}}(\bm{x},  \bm{Y}^{(N)})$ score in the EBM given by eq.~\eqref{EBM}. In addition, score matching approaches to training diffusion models can also be adopted with EBMs \cite{song2020score}, which further strengthens the connection between these models.

\section{Experiments}
\label{section-experiments}

\subsection{Experimental setup}
\subsubsection{Dataset}
The dataset used in this work is LJSpeech \cite{ljspeech17}, which includes 13,100 audio clips totalling approximately 24 hours from one female speaker. The dataset is split randomly into training (10,000 clips), validation (1800 clips) and test (1300 clips) sets. Objective evaluation is performed over the entire test set whilst subjective evaluation is performed over 100 randomly chosen test set clips. Front-end pre-processing of audio follows the open-source implementation available as a part of NVIDIA's Tacotron 2. \footnote{\scriptsize\url{https://github.com/NVIDIA/tacotron2}}

\subsubsection{Models}
The pre-trained TTS model providing hypothese for training and inference is Tacotron 2 \cite{shen2018natural}. The open-source implementation of NVIDIA using default configuration was adopted in this work. The structure of EBM follows the corresponding elements of Transformer-TTS \cite{li2019neural} available through an open-source implementation \footnote{\scriptsize\url{https://github.com/soobinseo/Transformer-TTS}} except that: 1) positional and character embeddings are 256-dimensional; 2) two EBMs with different dimensions of hidden features, 128 and 256 respectively, are explored in the study. Utterance-level energy is predicted by frame-level EBM energy prediction module, which consists of two 512-dimensional fully-connected layers.  Although it is possible to backpropagate gradients through the pre-trained TTS model, for simplicity this was not explored in this work. Both EBMs are trained for 125K iterations using Adam optimizer using batch size of 16 and a constant learning rate of $1 \times 10^{-4}$ on a single NVIDIA 3090 GPU. The number of parameters of these 2 EBMs and Tacotron 2 are shown in Table \ref{tab:parameters}. We use an open-source implementation\footnote{\scriptsize\url{https://github.com/NVIDIA/waveglow}} of the WaveGlow \cite{prenger2019waveglow} vocoder and adopt its default settings.

\subsubsection{Evaluation}
\label{sec:evaluation}
Mel cepstral distortion (MCD) 
, F0 frame error (FFE) and log-scale F0 root mean square error (log F0 RMSE)
 are adopted as objective metrics in this work. The MCD metric calculates distance between cepstral coefficient sequences of different lengths on the Mel frequency scale. The FFE metric measures discrepancy of fundamental frequency (F0) between synthesized and reference waveforms. Before objective calculating, dynamic time warping (DTW) is used to align the predicted mel-spectrogram and the reference. FAIRSEQ ${S^2}$ toolkit \cite{wang2021fairseq} is used to compute MCD and FFE scores. Mean opinion score (MOS) 
evaluation is conducted to evaluate speech naturalness by scoring each speech sample on a scale between 1 to 5 with 1 point intervals. Waveforms synthesized by 3 models compared in this work are mixed with test set waveforms. Each audio is listened to by 5 listeners, who are native English speakers, on the Amazon Mechanical Turk platform. 

\begin{table}[htbp]
  \centering
  \begin{tabular}{ l|ccc|c}
    \toprule
    \textbf{Model} & 
    \textbf{MCD $\downarrow$} & \textbf{FFE $\downarrow$} & $\log f_{\mathrm{o}} \downarrow$ & \textbf{Parameters} \\
    \midrule
     Tacotron 2 & 4.218 & 47.31\% & 0.292& 28.19M\\
     EBM$^{(1)}$ (small) & 4.163 & 47.06\% & 0.289& 2.30M\\
     EBM$^{(1)}$ (large) & 4.178& 47.05\% & 0.291& 7.64M\\
    \bottomrule
  \end{tabular}
  \caption{Comparison between Tacotron 2 and two EBMs utilising Tacotron 2 hypotheses as negative samples}
  \label{tab:parameters}
\end{table}

\subsection{Negative sampling methods}
Table \ref{tab:parameters} compares Tacotron 2 and two initial EBMs. These EBMs were trained using Tacotron 2 generated hypotheses as negative samples and a single step ($N=1$) Langevin MCMC, where $\mu$ was set to $0$ for simplicity and Adam rather than gradient descent update rule was adopted. Both large (256 hidden features) and small (128 hidden features) EBMs perform slightly better than the baseline.

Table \ref{tab:ablation} summarises performance of the alternative negative sampling methods (see Sec.~\ref{sec:training}) with the large EBM, where the simplified Langevin MCMC was run for $N=100$ steps. Many of these methods show significantly better performance than the baseline. Comparing between compressed and stretched spectral features suggest no strong preference for any particular method of time warping (TW). The method of 5\% time masking (TM) achieves lower MCD and FFE compared to other time masking methods, while the trend is opposite for frequency masking (FM), where higher percentage points (15\%) appear to be yielding better MCD results and worse FFE results. The likely reason is the 
negative interaction between FFE and frequency masking.
\begin{table}[htbp]
  \centering
  \begin{tabular}{ ll|cccc }
    \toprule
    \multicolumn{2}{l}{\textbf{Condition}} & \textbf{MCD $\downarrow$} & \textbf{FFE $\downarrow$} & $\log f_{\mathrm{o}} \downarrow$ \\
    \midrule
    \textbf{TM:} & 5\% & 4.149& 46.89\% & 0.290\\
       & 10\% & 4.166& 46.98\% & 0.284\\
       & 15\% & 4.161& 47.30\% & 0.285\\
    \midrule
    \textbf{FM:} & 5\% & 4.166& 46.88\% & 0.292\\
       & 10\% & 4.138& 47.27\% & 0.286\\
       & 15\% & 4.097& 47.35\% & 0.284\\
    \midrule
    \textbf{TW:} & 1.2 (compress) & 4.134 & 47.05\% & 0.291\\
       & 1.1 (compress) & 4.170 & 47.03\% & 0.291\\
       & 0.9 (stretch)& 4.168 & 47.28\% & 0.284\\
       & 0.8 (stretch) & 4.159 & 47.29\% & 0.286\\
     \midrule
    \textbf{RM:} & 25\% & $\bm {3.943}$ & $\bm {46.16\%}$ &0.282\\
    & 30\% & 4.013 & 46.57\% & $\bm {0.280}$\\
     \midrule
    & \textbf{Baseline} & 4.218& 47.31\%& 0.292& \\
    \bottomrule
  \end{tabular}
  \caption{Negative sampling methods (95\% confidence intervals)}
  \label{tab:ablation}
\end{table}

Table~\ref{tab:inference} investigates the impact of the Langevin MCMC steps on the performance of the best system in Table~\ref{tab:ablation}. 
\begin{table}[h]
  \centering
  \begin{tabular}{ r|ccc}
    \toprule
     \textbf{Step} & \textbf{MCD$\downarrow$} & \textbf{FFE$\downarrow$} &$\log f_{\mathrm{o}} \downarrow$\\
    \midrule
     0 & 4.218& 47.31\% & 0.292\\
      1 & 4.217& 47.31\% & 0.292\\
     100 & 3.943& 46.16\% & 0.282\\
     300 & $\bm {3.937}$& $\bm {45.85\%}$& $\bm {0.276} $\\
    \bottomrule
  \end{tabular}
    \caption{Simplified Langevin MCMC (95\% confidence intervals)}
  \label{tab:inference}
\end{table}
As the number of steps increases, the EBM applying 25\% random masking to negative samples performs better and better. 

Although random masking appears to be the most effective negative sampling method, the other masking methods may bring additional complementary information. Table~\ref{tab:combination} summarises performance of different combination approaches involving the RM 30\% EBM in Table~\ref{tab:ablation}. 
\begin{table}[htbp]
  \centering
  \begin{tabular}{ cccc|ccc }
    \toprule
    \textbf{RM} & \textbf{TM} & \textbf{FM} & \textbf{TW} & \multirow{2}*{\textbf{MCD $\downarrow$}} & \multirow{2}*{\textbf{FFE $\downarrow$}} & \multirow{2}*{$\log f_{\mathrm{o}} \downarrow$}\\
    30\%& 5\%& 5\%& 1.2& &\\
    \midrule
        $\checkmark$&&&& 4.013& 46.57\% & 0.280\\
        $\checkmark$& $\checkmark$&&& 3.958& 46.31\% & 0.276\\
        $\checkmark$&&$\checkmark$&& 3.997 & 45.63\% & 0.278\\
        $\checkmark$&&&$\checkmark$& 3.927 & 45.98\% & 0.267\\
        $\checkmark$&$\checkmark$&$\checkmark$&$\checkmark$& $\bm {3.882}$ & $\bm {45.36\%}$& $\bm {0.258}$\\
    \bottomrule
  \end{tabular}
  \caption{Combination of negative sampling methods (95\% confidence intervals)}
  \label{tab:combination}
\end{table}
All combinations examined perform better than using only random masking. The EBM making use of all masking methods performs the best.

\subsection{Subjective evaluation}
To solicit subjective assessment, a range of listening tests were conducted (see Sec.~\ref{sec:evaluation}). Table~\ref{tab:subjective_metrics} shows that the proposed EBM shows generally better MOS scores than the baseline Tacotron 2. 
\begin{table}[htbp]
  \centering
  \begin{tabular}{ l|c }
    \toprule
    \textbf{Method} & \textbf{MOS} \\
    \midrule
     Ground Truth & 4.53$\pm$0.05\\
     Ground Truth (log-Mel + WaveGlow) & 4.39$\pm$0.07\\
     Tacotron 2 (log-Mel + WaveGlow)& 3.77$\pm$0.11\\
     \midrule
     EBM (log-Mel + WaveGlow)& 3.84$\pm$0.13\\
    \bottomrule
  \end{tabular}
  \caption{Subjective evaluation}
  \label{tab:subjective_metrics}
\end{table}
Furthermore, the detailed breakdown of MOS score counts in Table~\ref{tab:MOS_count} shows that the EBM significantly reduced the number of MOS scores 2 (-2) and 3 (-27) and increase the number of MOS scores 4 (+26) and 5 (+3). 
\begin{table}[htbp]
  \centering
  \begin{tabular}{l|ccccc}
    \toprule
    \textbf{MOS} &\textbf{1} & \textbf{2} & 
    \textbf{3} & \textbf{4} & \textbf{5} \\
    \midrule
     \textbf{Tacotron 2} & 0& 3 & 121 & 344 & 15\\
     \textbf{EBM}  & 0& 1 & 94  & 370 & 18\\
    \bottomrule
    \end{tabular}
    \caption{MOS score counts}
  \label{tab:MOS_count}
\end{table}

\section{Conclusions}
\label{sec:conclusion}
This paper proposed a new class of non-autoregressive (non-AR) text-to-speech (TTS) models called energy-based models (EBM). As an example, it shows how powerful forms of EBMs can be designed by adopting architectures of state-of-the-art AR models like Transformer TTS. Although training models like EBMs is more complicated due to the intractability of normalisation terms, a range of training approaches is available. This paper describes how one such approach called noise contrastive estimation (NCE) can be adopted for training. As the NCE critically relies on the quality of negative samples used to contrast reference speech feature sequences, this paper proposed and evaluated a wide range of negative sampling methods. It found that random masking is the single best method but the combination of all proposed methods yielded the best performance. The paper also shows how sampling from EBMs can be performed by means of Langevin Markov Chain Monte-Carlo (MCMC). Since Langevin MCMC is closely linked with an iterative method used by popular diffusion models, the paper discusses similarities between EBMs and diffusion models. The paper concludes by subjective evaluation and finds that the proposed model provides improvements over Tacotron 2. Future work with EBMs will explore score parameterisation and the use of alternative TTS models for architectural and negative sampling choices. 

\bibliographystyle{IEEEbib}
\bibliography{strings,refs}

\end{document}